\documentstyle[11pt,twoside,psfig,jltp]{article}

\title{Nuclear Spin-Ordering Due to Correlated Atomic Motion in bcc $^3$He}

\author{N. Gov\address{Department of Physics,
University of Illinois at Urbana-Champaign, \\
1110 Green St., Urbana 61801, IL, USA} and E. Polturak \address{Physics Department,Technion-Israel Institute of Technology, Haifa 32000, Israel}}

\runninghead{N. Gov and E. Polturak}{Nuclear Spin-Ordering Due to Correlated Atomic Motion in bcc $^3$He}

\begin{document}

\begin{abstract}
We propose a new way to treat nuclear magnetism of solid $^{3}$He.
We argue that the magnetic interaction arises indirectly as a consequence of correlated
zero-point motion of the ions. This motion
lowers the energy of the ground state,
and results in a coherent state of oscillating electric dipoles. Distortion of the electronic
wavefunctions leads to hyperfine magnetic interactions with the nuclear spin.
 Our model describes
both the phonon spectra and the
 nuclear magnetic ordering of bcc $^{3}$He using a single parameter, the
 dipolar interaction energy $E_{0}$. The model yields correctly both the u2d2 symmetry
of the ordered phase and the volume dependence of the magnetic interaction.

\end{abstract}

\maketitle

     
\section{INTRODUCTION}

The spin-ordered phase of bcc $^{3}$He presents a difficult challenge to
accurate theoretical description \cite{fisher,cross,roger}. The main problem is to
explain why the transition temperature \cite{halperin} of 10$^{-3}$K is two
orders of magnitude larger than the nuclear dipolar interaction $\sim $10$%
^{-5}$K. 
We propose that magnetic
ordering is due to correlations in the zero point atomic motion. In this respect it is similar to the current Multiple-Spin Exchange (MSE) model \cite{roger}, though the zero-point atomic motion in our model
however does not involve the exchange of atoms \cite{niremil}. This model has conceptual problems that arise from the fact that in order 
to fit experimental data, many exchange cycles involving a large number of 
atoms are needed. In fact, it is not sure that this expansion converges \cite{cross,ceperley}, i.e. may not be the spin-ordering mechanism. Additionally, an overall consistent description of the 
experimental data has still not been achieved \cite{fisher,grey1}. It is 
therefore of interest to consider the possibility of another approach.

The zero-point
correlations we consider can be described as zero-point electric dipoles \cite{niremil}, and lower the energy of the ground state \cite{nirbcc}. This coherent state of oscillating electric dipoles modifies the transverse T$_{1}$(110) phonon
spectrum. We show here that this zero-point motion produces an oscillating magnetic polarization
of the electronic cloud which interacts with the nuclear spin. This hyper-fine type interaction has the right order of magnitude to be related to the spin-ordering transition, and
leads naturally to the distinct u2d2 antiferromagnetic phase.

\section{COHERENT DIPOLES AND TRANSVERSE PHONONS}

At temperatures which are high compared to the magnetic interactions (T$\gg $%
1mK), the local zero-point motion in bcc $^{3}$He can be treated in the same way as in $^{4}$He \cite{niremil}.
We begin by observing \cite{niremil,glyde} that the crystal potential is highly anisotropic in the bcc phase, being especially wide and anharmonic along the major axes (100,010,001). This means that the atomic wavefunctions will be particularly extended in these directions, resulting in dynamic-correlations, to reduce the overlap energy. The current treatment using variational wavefunction \cite{glyde} incorporates these correlations using a Jastrow-type function in a Self-Consistent Harmonic (SCH) calculation. This gives a satisfactory description of the phonon branches, except for the transeverse T$_{1}$(110) phonon, which in reality is much softer. Introducing cubic terms soften the phonon spectra, but now ruins the good agreement with the other branches.

We would like to describe the softening of this transverse phonon as resulting from the hybridization of the (virtual) harmonic phonon with a (virtual) local mode \cite{niremil}. The coupling is described as dipolar, where this zero-point dipole moment arises due to correlated atomic motion in the normal axes. The harmonic description of the crystal potential misses the low-lying vibration of the atoms due to the shallow (even double-well) potential in the normal directions, which is the one responsible for the dynamic correlations. We therefore treat them as two independent degrees of freedom. In this hybridization procedure we do rely on the SCH calculation, taken from previous works \cite{glyde}.

Directional oscillation of the nucleus will break the rotational symmetry of
the nuclear position relative to the electronic cloud. Going beyond the Born-Oppenheimer approximation, the energy due to relative nuclear-electronic fluctuations is \cite{vibronic}: $\Delta E \simeq (m/M)E_{sp} \simeq 10$K, where $m$ is the electron mass, $M$ the nuclear mass, and $E_{sp}$ the excitation energy of the He.
This is of the order of the energy involved in the zero-point mixing of the $s$ and $p$ electronic levels: $\left| \psi \right\rangle \simeq \left| s\right\rangle +\lambda \left| p\right\rangle $, i.e. $\lambda \sim \sqrt{\Delta E/E_{sp}}\sim 10^{-2}$. 
Since the nuclear motions are perfectly correlated this polarization results in zero-point dipolar interactions \cite{niremil}. The dipolar interaction energy $E_{0}$, of the order of $\Delta E$, is the energy associated with the zero-point correlated oscillations of the atoms along the major axes.

The ground-state of the
crystal in which the zero-point motion of the atoms is correlated may be
described as a global state of quantum resonance 
between the two degenerate configurations shown in Fig.1, each of
which minimizes the dipolar interaction energy \cite{niremil}.

\begin{figure}
\centerline{\psfig{file=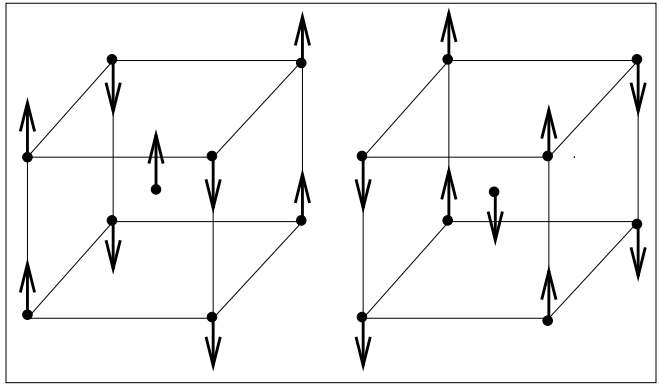,height=2.0in}}
\caption{The two degenerate 'antiferroelectric' dipole arrangement in the
ground-state of the bcc phase. The arrows show the instantaneous direction
of the electronic dipoles.}
\end{figure}

Due to the lower symmetry of
the dipolar array (Fig.1) compared to the crystal symmetry, a coupling between the
harmonic phonons and the dipolar modulation exists only along the (110)
direction \cite{niremil}. At the edge of the Brillouin zone
this transverse phonon has the energy $E_{0}$. In addition there appears a localized
excitation of energy $2E_{0}$, which is involved in mass diffusion and
contributes to the specific-heat \cite{niremil}, i.e. a quantum analogue of a
point defect.
There is good agreement of our approach with the experimental data for the T$_{1}$(110) spectrum for bcc $^4$He \cite{niremil}. This approch has the advantage over current techniques of explicitly describing the main quantum effect of the anisotropic and anharmonic potential. It also resolves previous inconsistencies relating to the properties of vacancies.

In bcc $^3$He, sound velocity data 
indicates that the slope of the T$_{1}$(110) phonon is about half of the SCH
calculation \cite{greywall,kohler}. We predict that this ratio should
indeed be 0.5, and therefore take half the energy of the calculated \cite{kohler} SCH 
T$_{1}$(110) phonon at the edge of the Brillouin zone to be the bare
dipole-flip energy $E_{0}\simeq 5$K at V=21.5cm$^{3}$/mole. According to our
model, the energy of the localized mode involved in thermally activated
self-diffusion is $2E_{0}\simeq 10$K. This value is in excellent agreement
with the activation energy measured by x-ray diffraction, ultrasonics and
NMR experiments \cite{heald} at V=21.5cm$^{3}$/mole. A similar activation
energy is also obtained from the excess specific heat \cite{grey}, and
pressure measurements \cite{izumi}. 
We therefore establish the likely occurance of coherent zero-point dipoles in the ground-state of bcc $^3$He.

\section{MAGNETIC INTERACTIONS}

\begin{figure}
\centerline{\psfig{file=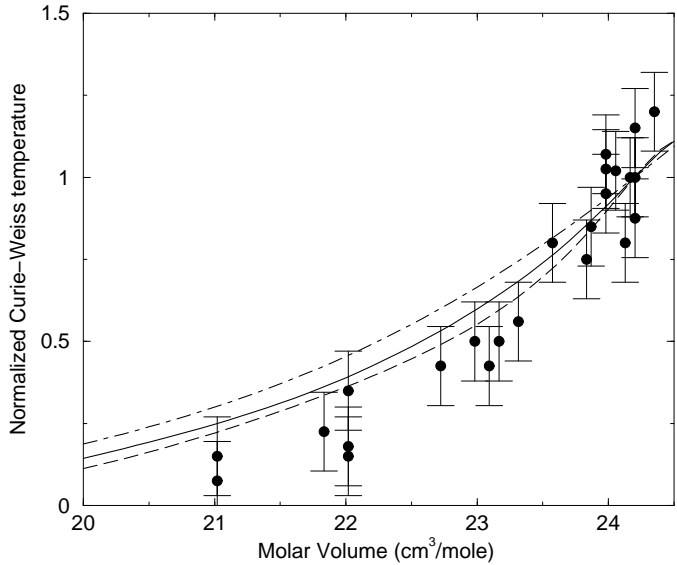,height=3.0in}}
\caption{The normalized magnetic splitting $2E_{mag}$ as a function of volume,
calculated using different experimental data sets for the activation energy $%
E_0$: dashed line [13], dash-dot [14] and solid line [15]. The solid circles are the normalized experimental Curie-Weiss temperature $\protect\theta$ [1].}
\end{figure}

We now describe the magnetic interaction arising from our model. The lowest $%
\left| p\right\rangle $ level of the He atom has the electrons in a spin $S=1
$ state due to strong exchange interaction \cite{cohen}, of the order of
0.25eV. In addition this level is split into 3 sublevels with $J=L+S=0,1,2$.
The splitting, due to spin-orbit coupling \cite{cohen}, is of the order of $%
\sim 1.5$K.\ In the ground state, the $\left| p\right\rangle $ electrons
will reside in the $^{3}P_{2}$ sublevel, with an oscillating magnetic moment
${\bf M}_{e}$ of size $\sim \lambda m_{e}$, where $m_{e}$ is the magnetic
moment of an electron. Because part of the magnetic moment is now in the $%
\left| p\right\rangle $ state, there appears a net uncanceled moment of
equal size in the $\left| s\right\rangle $ component of the electronic
wavefunction. 
In $^{3}$He the nuclear
magnetic spin $I=1/2$ will interact with the oscillating electronic magnetic
moment, mainly due to the contact term of the $\left| s\right\rangle $
electron at the nucleus. The magnetic interaction is of the hyper-fine type
\cite{cohen}, and the energy associated with it, $E_{mag}$, is given by:
$E_{mag}=\left\langle -\frac{8\pi }{3}{\bf M}_{e}\cdot {\bf M}_{n}\delta
(r)\right\rangle$, 
where ${\bf M}_{n}$ is the nuclear magnetic moment and the
calculation of the matrix element follows \cite{cohen}.
We show below that
the maximum value of $E_{mag}/k_{B}\simeq 0.75$mK (for V=24 cm$^{3}$/mole),
much larger than the direct nuclear dipole-dipole interaction, and is of the
right magnitude to explain the high transition temperature of nuclear
ordering in bcc $^{3}$He.

In our model, the
magnetic energy will change with pressure due to changes of
the electronic magnetic polarization ${\bf M}_{e}$. The main effect as the solid is compressed, is that the 3
sub-levels ($J=0,1,2$) broaden into partially overlapping bands.
The effect of the broadening of the
sub-levels with pressure can be approximated using the overlap integral of
the three sub-levels: 
$\left\langle {\bf M}_{e}(V)\right\rangle \simeq \left( 1-F(V)\right) \lambda
(V){\bf m}_{e}$, 
where the overlap integral $F(V)$
is calculated for the spin-orbit levels
using a simple band calculation. The level broadening is the Coulomb energy
of the $\left| p\right\rangle $ electrons due to overlapping wavefunctions on neighboring He nuclei \cite{ashcroft}.

We find that 
the overlap factor $1-F(V)$ is very
sensitive to volume, changing from 1 at V=24 cm$^{3}$/mole to $\sim 0.01$ at
V=19 cm$^{3}$/mole. One can see that as the volume decreases the broadening
of the bands increase, thereby decreasing the net magnetic polarization of
the oscillating electronic cloud.
The strength of the magnetic interaction should be proportional to the
measured Curie-Weiss temperature $\theta $. In Fig.2 we compare the normalized magnetic
splitting $2E_{mag}$ with the normalized values of the measured \cite{fisher} $\theta $. We find that the volume dependence of agrees very well.

We now consider the symmetry of the ordered spin system. The existence of
the hyper-fine splitting means that the simple quantum resonance condition
on each site is broken. Now, the two antiferoelectric configurations shown
in Fig. 1 are not degenerate, with an energy difference of $2E_{mag}$ per
site. It is possible to restore the degeneracy of the overall ground state,
and hence the quantum resonance condition. 
The possible arrangements of the nuclear spins that
fulfill the resonance requirement on each simple sublattice are those that
ensure an equal
number of atoms with electronic and nuclear spins aligned (and anti-aligned)
in both degenerate configurations of the electronic dipoles. These
arrangements preserve the overall time-reversal symmetry of the system at
zero field. We therefore end up
with an u2d2 arrangements which is the symmetry of the ordered nuclear phase
\cite{osheroff,bossy}. We point-out that the u2d2 phase results from symmetry considerations, independent of any quantitative parameters (as in MSE \cite{fisher,roger}).

\section{CONCLUSION}

To conclude, our model enables us to describe 
the nuclear magnetic ordering of bcc $^{3}$He using a single parameter,
which is the thermal activation energy $E_{0}$. Experimental values of $E_{0}$ were measured in several experiments. The model describes correctly both
the symmetry of the ordered phase and the volume dependence of the magnetic
interactions.

\section*{ACKNOWLEDGMENTS}
This work was supported by the Israel Science Foundation, by the Technion
VPR fund for the Promotion of Research,
the Fulbright Foreign Scholarship grant, 
the Center for Advanced Studies and NSF grant no. PHY-98-00978.

\end{document}